\documentclass[9pt, conference,a4paper]{APSIPA2025}
\usepackage{amsmath}
\usepackage{graphicx}
\usepackage{multirow}
\usepackage{threeparttable}
\usepackage[backend=biber,style=ieee]{biblatex}
\renewbibmacro{in:}{}  

\usepackage{subcaption}  
\usepackage{xcolor}
\usepackage{algorithm2e}
\usepackage{comment}
\usepackage{xcolor}
\usepackage{refcount}
\usepackage{amsfonts}

\usepackage{caption}
\usepackage{url} 
\usepackage{hyperref}

\usepackage{geometry}
\usepackage{fancyhdr}

\fancypagestyle{firststyle}{
  \fancyhf{}
  \fancyhead[C]{2025 Asia Pacific Signal and Information Processing Association Annual Summit and Conference (APSIPA ASC)}
}

\begin{document}

\title{Speaker Privacy and Security in the Big Data Era: \\ Protection and Defense against Deepfake}

\author{
\authorblockN{
Liping Chen\authorrefmark{1},
Kong Aik Lee\authorrefmark{2},
Zhen-Hua Ling\authorrefmark{1},
Xin Wang\authorrefmark{3},
Rohan Kumar Das\authorrefmark{4},
Tomoki Toda\authorrefmark{5},
Haizhou Li\authorrefmark{6}
}

\authorblockA{
\authorrefmark{1}
University of Science and Technology of China, China}

Emails: \{lipchen, zhling\}@ustc.edu.cn

\authorblockA{
\authorrefmark{2}
The Hong Kong Polytechnic University, China}
Email: kong-aik.lee@polyu.edu.hk

\authorblockA{
\authorrefmark{3}
National Institute of Informatics, Japan }
Email: wangxin@nii.ac.jp

\authorblockA{
\authorrefmark{4}
Fortemedia, Singapore }
Email: ecerohan@gmail.com

\authorblockA{
\authorrefmark{5}
Nagoya University, Japan}
Email:  tomoki@icts.nagoya-u.ac.jp

\authorblockA{
\authorrefmark{6}
School of Data Science, Chinese University of Hong Kong, China}
Email: haizhouli@cuhk.edu.cn

%
}

\maketitle
\thispagestyle{firststyle}
\pagestyle{fancy}

\begin{abstract}

In the era of big data, remarkable advancements have been achieved in personalized speech generation techniques that utilize speaker attributes, including voice and speaking style, to generate deepfake speech. This has also amplified global security risks from deepfake speech misuse, resulting in considerable societal costs worldwide. To address the security threats posed by deepfake speech, techniques have been developed focusing on both the protection of voice attributes and the defense against deepfake speech. Among them, the voice anonymization technique has been developed to protect voice attributes from extraction for deepfake generation, while deepfake detection and watermarking have been utilized to defend against the misuse of deepfake speech. This paper provides a short and concise overview of the three techniques, describing the methodologies, advancements, and challenges. A comprehensive version, offering additional discussions, will be published in the near future.

\end{abstract}

\let\thefootnote\relax\footnotetext{This work was supported in part by the National Key Research and Development Program of China (Project No. 2024YFE0217200) and the Innovation and Technology Fund of the Hong Kong SAR (Project No. MHP/048/24), and JST PRESTO (Project No. JPMJPR23P9).}


\section{Introduction}
In the past years, the evolution of the Internet has enabled people to generate and share an increasing amount of speech data online. Simultaneously, advancements in computational power have significantly enhanced the development of deep neural networks (DNNs), providing effective tools for processing the information conveyed by big speech data, resulting in substantial progress in speech technology \cite{li2022recent,kumar2023deep,bai2021speaker}. As a crucial component of the information conveyed by speech signals, modeling techniques for speaker attributes have gone through remarkable advancements, prompting the development of related speech techniques. For example, given a speech segment of only a few seconds, the speaker's identity can be recognized using speaker recognition techniques. Moreover, it can also be utilized to synthesize the speaker's speech using personalized speech generation techniques, generating deepfake speech of the speaker. Notably, the emergence and rapid development of large speech generation models \cite{chen2025neural,du2024cosyvoice,megatts3,FACodec} have greatly enhanced the fidelity and speaker similarity regarding voice and speaking style of the generated speech.

Given that voice and speaking style are unique to each individual, thereby conveying speaker privacy information, the misuse of deepfake speech poses significant security threats globally, as extensively reported in media sources. For instance, an individual's deepfake speech may be exploited for fraudulent activities or disseminated to harm her/his reputation. When public figures are targeted, deepfake speech becomes a powerful tool for manipulating public sentiment and opinions. In the big data era, the security challenges posed by deepfakes have drawn governmental attention worldwide, driving the formulation of regulatory frameworks, such as the General Data Protection Regulation in European Union \cite{gdpr}, the Interim Regulations for the Management of Generative Artificial Intelligence Services in China \cite{CNLaws}, the Act on the Protection of Personal Information in Japan \cite{JPLaws} and the Personal Data Protection Act in Singapore \cite{SGLaws}.

The security threats posed by deepfake speech have also attracted the attention of the research community, prompting the development of protection and defense techniques against it. Specifically, the protection techniques aim to protect speaker attributes from being extracted and exploited for deepfake generation, among which voice anonymization provides a viable solution \cite{tomashenko2020introducing}. The defense techniques are developed to prevent the use of deepfake speech for malicious purposes, wherein both deepfake detection~\cite{yi_audio_2023} and watermarking~\cite{reilly2025deep} are among the viable techniques. Deepfake detection operates in a passive manner, without prior knowledge of the speech synthesis system; whereas watermarking can function proactively, as the inaudible watermark is embedded into the generated speech data before it is distributed to the public.

\begin{figure*}[t]
    \centering
    \includegraphics[scale=0.58]{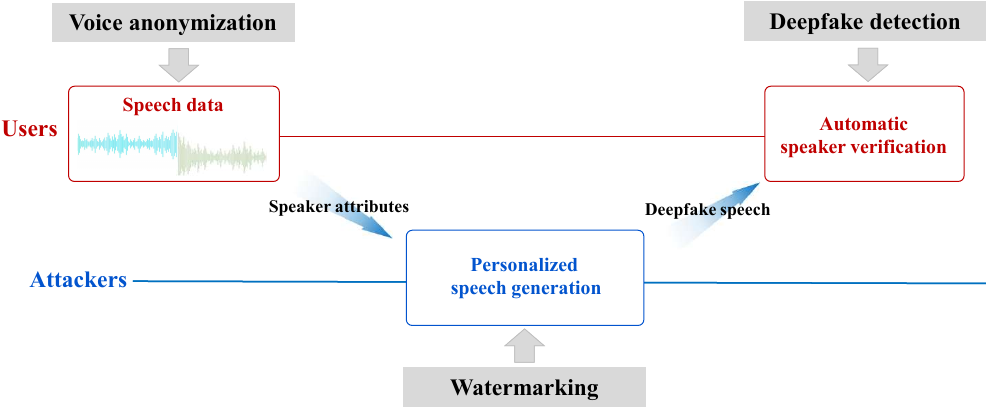}
    \caption{Illustrative example of applications for voice anonymization, deepfake detection, and watermarking techniques, shown in the grey boxes, for the security of speaker privacy. Voice anonymization and deepfake detection are utilized at the user end, while watermarking is applied during the generation of deepfake speech.}
\label{fig: diagram}
\end{figure*}

Fig. \ref{fig: diagram} illustrates an example for the applications of the three techniques for the security of speaker privacy against deepfake speech. In the absence of security techniques, leveraging the speaker attributes conveyed in the speech utterance of a speaker, deepfake speech can be generated by the attacker using personalized speech generation techniques that replicate the speaker's voice and mimic her/his speaking style. The deepfake speech can then be maliciously utilized to impersonate the speaker and attack automatic speaker verification (ASV) systems. To prevent the extraction of voice attributes for deepfake speech generation, voice anonymization techniques can be utilized to obscure or eliminate these attributes in the speaker's utterance, thereby preventing the synthesis of speech in their voice. Additionally, deepfake detection can be employed before the ASV system to examine whether the input speech is synthetic. Furthermore, by integrating watermarking techniques into the personalized speech generation process, deepfake speech can be marked as synthetic rather than genuine. Before the ASV system, the watermark can be detected, thereby further inhibiting its potential use for impersonation. This paper provides a concise overview of the three techniques, outlining the methodologies, advancements, and challenges.


\section{Voice anonymization}
Voice anonymization is a technique that can be dated back to the 1980s, when speech signals were represented as analog signals \cite{analogVPS,jayant1983analog,analog_voice_anonymization_TFSP}. At that time, the speech signals were modified to protect the speaker attributes within them, resulting in degraded speech quality. In recent years, advancements in neural networks and speech generation techniques have significantly enhanced the development of voice anonymization \cite{voiceprivacy, fang2019speaker}. As voice attributes are detectable by both human auditory systems and machine algorithms, voice anonymization can be realized in both synchronous \cite{Vaidya,SpeechSanitizer,McAdams,voicemask,Hidebehind, 10945923} or asynchronous \cite{chen2023voicecloak,UniAP,chen2024adversarial,MitigatingTTS,any2any,wang2024async,wangSPL} manners. Synchronous anonymization protects voice attributes from being correctly perceived by both human hearing and machine algorithms, whereas asynchronous anonymization preserves human perception and solely protects against machine algorithm extraction.

The requirements for anonymized speech are defined from two aspects: voice privacy protection capability and utility \cite{tomashenko2020introducing,vpc2024}. Voice privacy protection requires that the voice attributes can not be correctly extracted and utilized by machine algorithms. In terms of human perception, synchronous voice anonymization is required to alter the original voice attributes, while asynchronous voice anonymization needs to preserve them as perceived by human listeners. The utility depends on the applications and may require that anonymized speech preserves the non-voice attributes of the original speech, including the speech quality, linguistic content, and prosody, among others.

This section outlines the generative and adversarial approaches to voice anonymization, which currently attract the most attention in the research community. Specifically, the generative approach produces anonymized speech as synthesized output, applicable for both synchronous and asynchronous voice anonymization. The adversarial approach obtains anonymized speech by adding perturbations to the original signal, facilitating asynchronous voice anonymization.

\subsection{Generative voice anonymization}
Based on a speech generation framework that disentangles and represents speaker attributes as embedding vectors, voice anonymization can be accomplished by substituting the original speaker embedding with that of a pseudo-speaker. This mechanism facilitates generative solutions for both synchronous and asynchronous voice anonymization, wherein the anonymized output is synthesized speech. In synchronous voice anonymization, the pseudo-speaker embedding is obtained as a representation of a speaker that is distinct from the original speaker \cite{fang2019speaker,cohort-speaker,psd}. As reported in the VoicePrivacy Challenge 2024 \cite{vpc2024_summary_slides}, the voice anonymization approaches demonstrated effectiveness in protecting speaker privacy while preserving linguistic content. As voice anonymization focuses on protecting the voice attribute of speakers, the disentanglement of voice attribute from the others forms the crux of the generative methodology. Due to the correlation between voice and prosody attributes, both encoded in prosodic features like pitch and energy, preserving original prosody while preventing speaker attribute leakage through these features presents a significant challenge.

In asynchronous voice anonymization, limited by the requirement to preserve human perception of the original speaker, the pseudo-speaker embedding is obtained through constrained modifications of the original one. Existing studies have shown that machine recognition of original speaker attributes can be effectively obscured in asynchronously anonymized speech, with their human perception preserved \cite{wang2024async,wangSPL}. However, compared to synchronous voice anonymization, additional challenges exist including: 1) due to the lack of disentanglement of machine and human-perceived attributes within the speaker embedding, the modifications on speaker embedding have to reach trade-offs between the obscuration of machine perception with the preservation of human perception; 2) in scenarios where attackers can access the anonymization system to train speaker attribute extractors, the effectiveness in voice protection drops significantly.

\subsection{Speaker-adversarial speech}
The speaker-adversarial speech provides an alternative solution to asynchronous voice anonymization. As discovered in \cite{2014Explaining}, neural network models were susceptible to adversarial perturbations in input samples \cite{2014Explaining}, leading to investigations into adversarial attacks on speaker recognition models \cite{zhang2023imperceptible, abdullah2021hear, 9053076, 9053076}. The ability of speaker-adversarial speech to deceive speaker recognition further enables its application in voice anonymization \cite{chen2023voicecloak,UniAP,chen2024adversarial,MitigatingTTS,any2any}. Existing studies have demonstrated that speaker-adversarial speech effectively prevents the accurate extraction of voice attributes by the speaker extractor that is used for adversarial perturbation generation. Besides, its efficacy in preserving the human perception of speaker attributes and the utility of the original speech has been validated. However, in its application in voice anonymization, the technique faces challenges including: 1) the limited transferability of the perturbation leads to a substantial reduction in its capability of preventing the voice attributes from being extracted by speaker extractors that were not involved in the adversarial perturbation generation process, 2) similar to the generative approach, speaker-adversarial speech loses its voice privacy protection effectiveness if an attacker gains access to its generation system and uses the adversarially generated speech to train a speaker attribute extractor.


Speaker privacy covers beyond voice to include speaking style, lexical preferences, and grammatical choices; therefore, voice anonymization alone is insufficient to fully protect speaker privacy, necessitating more complex approaches than these mentioned methods.

\section{Deepfake detection}
Deepfake detection processes an input utterance and yields an answer indicating whether the utterance was uttered by a human speaker or not. While the formulation is not flawless, deepfake detection as a binary classification task has received growing attention in the past decade~\cite{yi_audio_2023, li_survey_2025}. Its assumed application includes what Fig.~\ref{fig: diagram} illustrates -- the deepfake detector rejects any input trial that is unlikely to be a human speech before the trial is delivered to an ASV system.~
In that context, `deepfake detection' was often referred to as speech anti-spoofing~\cite{wu2015asvspoof}.
Another application is to protect the human ears, for example, by tagging the sound track of a video as deepfake on social media platforms. It is in the second application that we see an increased number of incidents caused by high-quality deepfake speech contents~\cite{mcafee_beware_2023}. In the rest of this section, however, we use the term deepfake detection to cover both applications.

\subsection{Progress of deepfake detection}
Approaching deepfake detection as a binary classification task allows us to plug various modules into the machine learning pipeline, which accelerate the research and development iteration in the field. 
Feature engineering has advanced from purely using digital signal processing (DSP) algorithms~\cite{kamble_advances_2020} (e.g., linear-frequency cepstrum coefficients ~\cite{sahidullah2015comparison}) to the hybrid of DSP and deep learning methods (e.g., trainable filters integrated in a convolution network~\cite{tak2020end}). The latest trend is to extract features using pre-trained self-supervised-learning (SSL)-based models~\cite{mohamedSelfSupervised2022}. 
Potentially due to the SSL training based on a large-scale of human speech data, the SSL-based models seem to be able to extract features that better reveal the artifacts in the frequency band of speech sounds~\cite{wang_investigating_2022, tak_automatic_2022}. 

On the classifiers, we have witnessed the paradigm shift from linear models (e.g., Gaussian mixture model) to various types of DNNs. Many of them are borrowed from computer vision or deep learning fields, e.g., light convolutional neural network (CNN)~\cite{lavrentyeva19_interspeech}, ResNet~\cite{zeinali19_interspeech}, graph-based network~\cite{jung2022aasist}, and the latest state-space models~\cite{xiao_xlsr-mamba_2025}. Combining the SSL-based feature extractor and the latest DNN-based classifier appears to be the state-of the-art paradigm~\cite{li_survey_2025}.


The progress of deepfake detection is also supported by databases. Notable databases are those from the ASVspoof challenges~\cite{
wang2024asvspoof5} and audio deepfake detection challenges~\cite{ yi_add_2023}. It is based on the shared databases (as well as evaluation protocols) that results from different papers are compared.
Nowadays, many research groups are creating new databases that span different languages~\cite{muller_mlaad_2024}, newest fake speech generation methods~\cite{muller_mlaad_2024, lu24f_interspeech}, more challenge acoustic conditions~\cite{jung_spoofceleb_2025}, all of which address new research questions. 

\subsection{Challenges and future directions}
High-performing detectors can now attain significantly low equal error rates on evaluation datasets with limited complexity, such as ASVspoof 2019.
From this paragraph, however, we dive into the potential limitations and issues.

\subsubsection{Short-cut learning}
Classifiers are prone to short-cut learning, and binary deepfake detectors are no exception.
Specifically, a deepfake detector may overfit to the artifact that only presents in a particular training set but is irrelevant to the decision~\cite{sahidullah_shortcut_2025} (e.g., the length of the non-speech region~\cite{muller21_asvspoof}). Although the detector may perform well on a test set with a similar artifact, it cannot make useful decisions when the short-cut does not exist in a different test set. Even worse, the detector can be easily fooled by an attacker who adversarially uses the artifact to mislead the detector. Avoiding short-cut learning in model training is a good research direction. 
For studies not directly addressing the short-cut learning issue, we call for evaluation using multiple test sets from different data sources (e.g., those from the ASVspoof plus other non-ASVspoof test sets~\cite{speecharena-df-leaderboard}). This avoids over-optimistic results caused by short-cut learning on a particular database.

\subsubsection{Explainability}
Modern detectors are predominantly black-box models that output decision scores without revealing the underlying reasoning. This does not help when evidence for the decision is required, for example, by the system user. 
Some studies try to interpret the detectors' behaviors using simpler models, but they are unlikely to be logically solid~\cite{Rudin2019}.\footnote{Simpler models approximate the black-box detector's decision in a neighborhood of the feature space. If the approximation is sufficiently accurate, we should use the simpler models rather than the black-box detector; if the approximation is inaccurate, the interpretation does not make sense.} 
A few recent studies explored explainable models by design~\cite{mishra_towards_2026}. 
There is also an effort of using audio large language model to produce text-based explanation~\cite{gu_allm4add_2025}. These are encouraging directions that are indispensable for decisions making with evidence.

\subsubsection{Task definition}
While a binary classification task is easy to work with, the definition of the two classes is equivocal. 
Following the convention, we may start by defining the negative class as `being generated by a model given a text or a source voice prompt'. Then the positive class will be simply the logic negation of the negative. However, should a human-uttered utterance be treated as `fake' if it is compressed using a DNN-based speech codec? Note that the DNN-based decoder in many speech codec share the same DNN architecture (e.g., HiFiGAN~\cite{NEURIPS2020_c5d73680}) as many speech synthesis systems. 
As another example, should an anonymized utterance be tagged as fake?


What make things worse is that a binary deepfake detector is also expected to generalizable to any unseen attack. 
However, this goal may be untestable --- we cannot claim a deepfake detector to be generalizable to \emph{any} unseen fake speech unless we evaluate on \emph{all} possible fake speech, but there will always be new fake speech created in the future that is not test. 

A safer approach may be to define the two classes based on the applications, 
but the detector has to be rebuilt whenever the class definition is updated. 
Alternatively, we may shift from the binary classification task to different paradigms, e.g., spoofing-oriented verification~\cite{chen2021spoofprint} and multi-class source tracing~\cite{klein2024source, muller24_interspeech}. We may also hide information into speech data to signify the source (either human or a particular generator). Then the deepfake detection becomes a watermarking task that will be explained in the next section.

Practitioners working on standard databases may not bother with the above issues, but they are unavoidable when deploying deepfake detectors in real applications.
Hence, we encourage the readers to keep an eye on potential issues. 

\section{Watermarking}
Watermarking involves embedding imperceptible messages into a data sample and extracting this message at a later stage~\cite{coxSecure1996}. The technique was initially applied to multimedia data for copyright protection, authentication, digital ownership management, among others~\cite{coxSecure1996}, but it  recently found application in proactive image and speech deepfake detection. 
For example, watermark can be embedded into synthetic speech~\cite{liu2023detecting, chen2023wavmark}, indicating that the speech is generated rather than authentic, thereby providing a viable solution for preventing the misuse of both voice and speaking style information of the original speaker. 
While the requirements may vary among the applications, in most of cases watermark for speech data must be imperceptible and robust.
An imperceptible watermark message should not degrade the perceptual quality of the carrier speech data. A robust watermark should persist even if the carrier speech signal is subjected to intentional manipulation or unintentional degradation, for example, MP3 compression and low-pass filtering.\footnote{The terms `intentional' and `unintentional' are defined from the attacker's perspective. For example, an attacker may intentionally use MP3 compression to destroy the watermark; degradation may also be unintentional if the watermarked data has to be compressed using MP3 when being transmitted.}

\subsection{Post-processing and collaborative approaches}
Speech watermark using classical signal-processing-based methods have been extensively investigated before the deep learning and big data era~\cite{hua2016twenty}. Despite well-crafted algorithms and thorough theoretical analysis~\cite{cox1999watermarking} (with certain assumptions on channel noise), the classical methods seem to be unable to produce watermark robust to many types of degradation~\cite{yamniEfficient2023, chen2023wavmark, wenSoK2025, reilly2025deep}. Hence, some recent studies started to investigate deep-learning-based speech watermark.

Popular designs can be roughly categorized into post-processing and collaborative types. The former takes speech data (either synthetic or real) as system input and add watermark in a post-processing manner~\cite{chen2023wavmark, liu2023detecting, o2024maskmark, roman2024proactive, wuAudio2024}. Most of these methods use an encoder-decoder-like DNN --- the watermark bit string is added to the output of the speech encoder, and the watermarked speech waveform is reconstructed by the decoder. There is also a method adding learnable perturbation to the input speech waveform~\cite{wuAudio2024}. The post-processing approaches can be easily applied to any speech synthesis system, and many of them demonstrated higher robustness than signal-processing-based ones~\cite{chen2023wavmark, wenSoK2025}.

The collaborative approach focuses on watermarking the speech generation model~\cite{juvela24-collaborative-watermarking, juvela2024audio, cheng2024hifi, zhaoTraceable2025a}. 
The generation model and a watermark detector are jointly optimized so that the generated speech carries imperceptible latent information that can be extracted by the watermark detector (but not any other detector).
This is different from the post-processing approach, wherein the generation model is independent from the watermarking algorithm. Although the joint optimization complicates the training procedure, the collaborative approach seems to be more robust against many degradation types than the post-processing methods~\cite{reilly2025deep}.

\subsection{Challenges and future directions}
Unfortunately, watermark produced by the recent DNN-based algorithms are not sufficiently robust. 
Among these, processing with a vocoder or codec results in strong distortion of the watermark information~\cite{liu2024audiomarkbench}.
The DNN-based codec is particularly `effective' in removing the watermark~\cite{reilly2025deep, ozerComprehensive2025}. The lack of general robustness is problematic if the watermarked data is distorted by similar vocoder or codec during the transmission or under attacker. 
Besides, the defense against adversarial attacks on the watermark detectors remains an open question that awaits future investigation. 

Even being sufficient robust against unintentional or intentional degradation, the current speech watermark algorithms do not guarantee security~\cite{cayreWatermarking2005}. 
For example, in application where watermarks are added to real and synthetic speech data, an attacker may add another layer of conflicting watermark (e.g., adding a `real' watermark to a synthetic utterance that already carries the `fake' watermark). This collusion attack is beyond the issue of robustness against vocoder or codec and may not be easily fixed by optimizing the watermark encoder and decoder. Instead, protocols of embedding and authenticating the watermark may be necessary~\cite{steinebach2003watermarking, cayreWatermarking2005}.

\section{Conclusions}
This paper provides a concise overview of techniques employed to address the security threats caused by deepfake speech, including voice anonymization, deepfake detection, and watermarking. Their methodologies, advancements, and challenges are discussed. Moreover, potential challenges may arise in integrating the three techniques into a unified system, necessitating further investigation and not addressed in this paper.


\printbibliography
\end{document}